\documentstyle[aps,psfig,prb]{revtex}
\begin{document}
\baselineskip20pt
\thispagestyle{empty}
\pagestyle{plain}
\title{
Theory of magnetic order in the three-dimensional spatially
anisotropic Heisenberg model}
\author{L.\ Siurakshina \cite{ljuda} and D.\ Ihle}
\address{Institut f\"ur Theoretische Physik, Universit\"at Leipzig, 
D-04109 Leipzig, Germany}
\author{R.\ Hayn}
\address{Institut f\"ur Theoretische Physik, Technische Universit\"at Dresden,
D-01062 Dresden, Germany}
\date{\today}
\maketitle

\begin{abstract}
A spin-rotation-invariant Green's-function theory of long-range and
short-range order (SRO) in the $S=1/2$ antiferromagnetic Heisenberg model with
spatially anisotropic couplings on a simple cubic lattice is presented. The
staggered magnetization, the two-spin correlation functions, the correlation
lengths, and the static spin susceptibility are calculated self-consistently
over the whole temperature region, where the effects of spatial anisotropy
are explored. As compared with previous spin-wave approaches, the N\'{e}el
temperature is reduced by the improved description of SRO. The maximum in the
temperature dependence of the uniform static susceptibility is shifted with
anisotropy and is ascribed to the decrease of SRO with increasing
temperature. Comparing the theory with experimental data for the magnetization
and correlation length of La$_2$CuO$_4$, a good agreement in the temperature
dependences is obtained.
\end{abstract}


\section{Introduction}
The magnetic properties of spatially anisotropic antiferromagnetic (AFM)
quantum spin systems, such as the quasi-two-dimensional (2D) parent compounds
of high-$T_c$ superconductors (e.\ g.\ La$_2$CuO$_4$, Ca(Sr)CuO$_2$)
\cite{Kam94}  and the  
quasi-1D cuprates Sr$_2$CuO$_3$, Ca$_2$CuO$_3$, \cite{Koj97,REH97} and
SrCuO$_2$, \cite{ZBK99} are of current interest. The main problem is the
influence of spatial anisotropy on the staggered magnetization $m$ and the
Ne\'{e}l temperature $T_N$ in the 3D spin-$\frac 12$ AFM Heisenberg model
\begin{equation}
H = J_x \left[
\sum_{\langle i j \rangle_x} {\bf S}_i {\bf S}_j + 
R_y \sum_{\langle i j \rangle_y} {\bf S}_i {\bf S}_j + 
R_z \sum_{\langle i j \rangle_z} {\bf S}_i {\bf S}_j \right]
\; . 
\label{1}
\end{equation}
Here $R_y=J_y/J_x$, $R_z=J_z/J_x$ (throughout we set $J_x=1$) and $\langle i j
\rangle_{x,y,z}$ denote nearest-neighbor (NN) bonds along the $x$-, $y$- or
$z$-directions of a simple cubic lattice. For real systems, we consider $0 \le
R_z \ll R_y \le 1$.

In the paramagnetic phase, there exists a pronounced AFM short-range order
(SRO) which is reflected by a maximum in the temperature dependence of the
magnetic susceptibility at $T_{max}$, where
$0.64<T_{max}<1.2$. \cite{JohR} However, the RPA spin-wave theories
\cite{MSS92,STS92} and the mean-field theories using auxiliary-field
representations (Schwinger-boson, \cite{Kop92,Kei92} Holstein-Primakoff, 
\cite{Liu90,MMR97} Dyson-Maleev, \cite{Liu90,SI93} and boson-fermion
representations \cite{IKK99}) which were developed for the quasi-2D model with
$R_y=1$, are valid only at sufficiently low temperatures. 
In those theories, the temperature dependent SRO is not adequately taken into
account; in particular, the maximum in the magnetic susceptibility cannot be
reproduced. In the chain mean-field approaches, \cite{Sch96} recently
improved by spin-fluctuation corrections \cite{IK99} which lower $m$ and
$T_N$, an asymmetry between intrachain and interchain correlations is
introduced. As was shown in Ref.\ \onlinecite{REH97} on the basis of a
detailed estimate of the exchange integrals for the quasi-1D cuprates using a
first-principle calculation 
(Sr$_2$CuO$_3$: $R_y \simeq 0.004$; Ca$_2$CuO$_3$: $R_y \simeq 0.02$), all
previous approaches overestimate both $m$ and $T_N$. This 
deficiency is calling for a theory that provides an improved description of
SRO over the whole temperature region. In Ref.\ \onlinecite{WI97}, a
spin-rotation-invariant Green's-function theory for the 2D isotropic
Heisenberg and $t$-$J$ models was developed, which yields a good description
of spin correlation functions of arbitrary range and at arbitrary
temperatures. Moreover, the susceptibility maximum was obtained in good
agreement with quantum Monte Carlo calculations. Applying this
approach to the 2D anisotropic Heisenberg model, \cite{ISW99} the short-ranged
spin correlations at $T=0$ are well reproduced as compared with exact
diagonalization (ED) data. Accordingly, we expect such a theory to describe
the SRO 
properties quite well also in the 3D model (\ref{1}).

In this paper we extend the Green's-function approach of Refs.\
\onlinecite{WI97} and \onlinecite{ISW99} and present a theory of AFM
long-range order (LRO) and SRO for the 3D anisotropic Heisenberg model
(\ref{1}) (Sec.\ II). Thereby, the correlations along all spatial directions
are described on the same footing. In Sec.\ III the ground state is
investigated, where the magnetization and short-ranged spin correlation
functions are calculated. In Sec.\ IV we present our finite-temperature
results on the $R_z$-dependence of $T_N$, $m(T)$, and of the AFM correlation 
lengths. Moreover, for the first time, the effects of an arbitrary spatial
anisotropy on the temperature dependence of the uniform static spin
susceptibility, especially on $T_{max}$, are investigated. The results are
compared with experiments on La$_2$CuO$_4$ (magnetization, correlation length,
magnetic susceptibility). The summary of our work can be found in Sec.\ V.

\section{Dynamic spin susceptibility}

To determine the dynamic spin susceptibility $\chi^{+-}({\bf q},\omega)=
- \langle\langle S_{{\bf q}}^+;S_{-{\bf q}}^- \rangle\rangle_{\omega}$ by the
projection 
method outlined in Ref.~\ \onlinecite{WI97}, we choose the two-operator basis
${\bf A} = \left( S_{{\bf q}}^+, i \dot{S}_{{\bf q}}^+ \right)^T$ 
and consider the two-time retarded 
matrix Green's function in a generalized mean-field approximation, 
$\langle\langle {\bf A};{\bf A}^+\rangle\rangle_{\omega} 
= \left[\omega - {\bf M}^{\prime}
{\bf M}^{-1}\right]^{-1} {\bf M}$ with ${\bf M} 
= \langle [ {\bf A},{\bf A}^+]\rangle$ and 
${\bf M}^{\prime} = \langle [ i {\bf \dot{A}},{\bf A}^+ ] \rangle$, using
Zubarev's notation. \cite{Zu60} 
We get 
\begin{equation}
\chi^{+-}({\bf q},\omega)=
-\frac{M_{{\bf q}}^{(1)}}{\omega^2-\omega_{{\bf q}}^2} 
\; . 
\label{2}
\end{equation}
The spectral moment $M_{{\bf q}}^{(1)}=\langle \left[ i \dot{S}_{{\bf q}}^+,
S_{-{\bf q}}^- \right] 
\rangle$ is given by 
\begin{equation}
M_{{\bf q}}^{(1)}=-4C_{1,0,0} (1-\cos q_x) - 4 R_y C_{0,1,0} (1-\cos q_y) - 4
R_z 
C_{0,0,1} (1 - \cos q_z)
\; . 
\label{3}
\end{equation}
The two-spin correlation functions $C_{{\bf r}} = \langle S_0^+ S_{{\bf r}}^-
\rangle \equiv C_{n,m,l}$ with ${\bf r}=n {\bf e}_x + m {\bf e}_y + l
{\bf e}_z$ are calculated from
\begin{equation}
C_{{\bf r}}=\frac{1}{N} \sum_{{\bf q}} C_{{\bf q}} \mbox{e}^{ i 
{\bf q}{\bf r}} 
\quad , \quad
C_{{\bf q}}=\frac{M_{{\bf q}}^{(1)}}{2 \omega_{{\bf q}}}
\left[ 1 + 2 n(\omega_{{\bf q}}) \right] \; , 
\label{4}
\end{equation}
where $n(\omega_{{\bf q}})=\left( \mbox{e}^{\omega_{{\bf q}}/T} - 1
\right)^{-1}$. The NN correlation functions are directly related to the
internal energy per site by $\epsilon = \frac 32 \left( C_{1,0,0} + R_y
C_{0,1,0} + R_z C_{0,0,1} \right) $. 

To obtain the spectrum in the approximation $- \ddot{S}_{{\bf q}}^+ =
\omega_{{\bf q}}^2 S_{{\bf q}}^+$, we take the site representation and
decouple the products of three spin operators in $-\ddot{S}_i^+$ along NN
sequences introducing vertex parameters in the spirit of the scheme proposed
by Shimahara and Takada \cite{ST91} and extending the decoupling given in
Ref.\ \onlinecite{ISW99},
\begin{equation}
S_i^+ S_j^+ S_l^- = \alpha_1^{x,y,z} \langle S_j^+ S_l^- \rangle S_i^+ 
+ \alpha_2 \langle S_i^+ S_l^- \rangle S_j^+ \; .
\label{5}
\end{equation}
Here $\alpha_1^x$, $\alpha_1^y$, and $\alpha_1^z$ are attached to NN
correlation functions along the $x$-, $y$- and $z$-directions, respectively,
and $\alpha_2$ is associated with the longer ranged correlation functions. We
obtain
\begin{eqnarray}
\omega_{{\bf q}}^2 &=& 1+R_y^2+R_z^2-\cos q_x -R_y^2\cos q_y - R_z^2\cos q_z 
+ 2 \alpha^x_1C_{1,0,0}\cos(2q_x)-2\alpha_2C_{2,0,0}\cos q_x
\nonumber\\ 
&&+ 2R_y^2 ( \alpha^y_1C_{0,1,0} \cos(2q_y)-\alpha_2C_{0,2,0}\cos q_y)
+ 2 R_z^2 ( \alpha_1^zC_{0,0,1}\cos(2q_z)-\alpha_2C_{0,0,2}\cos q_z)
\nonumber\\
&&- 2\alpha_1^xC_{1,0,0} \cos q_x +2\alpha_2C_{2,0,0}
- 2R_y^2 ( \alpha_1^y C_{0,1,0} \cos q_y -\alpha_2 C_{0,2,0})
- 2 R_z^2 ( \alpha_1^z C_{0,0,1} \cos q_z -\alpha_2 C_{0,0,2})
\nonumber\\
&&+ 4 R_y (( \alpha_1^x C_{1,0,0} + \alpha_1^y C_{0,1,0}) \cos q_x \cos q_y 
- \alpha_2 C_{1,1,0} ( \cos q_x + \cos q_y ))
\nonumber\\
&&+ 4 R_z ((\alpha_1^x C_{1,0,0} + \alpha_1^z C_{0,0,1}) \cos q_x \cos q_z 
- \alpha_2 C_{1,0,1} ( \cos q_x + \cos q_z ))
\nonumber\\
&&+ 4 R_y R_z (( \alpha_1^y C_{0,1,0} 
+ \alpha_1^z C_{0,0,1}) \cos q_y \cos q_z 
- \alpha_2 C_{0,1,1} ( \cos q_y + \cos q_z ))
\nonumber\\
&&- 4 R_y (\alpha_1^x C_{1,0,0} \cos q_y  + \alpha_1^y C_{0,1,0} \cos q_x 
- 2 \alpha_2 C_{1,1,0})
- 4 R_z (\alpha_1^x C_{1,0,0} \cos q_z  + \alpha_1^z C_{0,0,1} \cos q_x 
- 2 \alpha_2 C_{1,0,1})
\nonumber\\
&&- 4 R_y R_z (\alpha_1^y C_{0,1,0} \cos q_z  + \alpha_1^z C_{0,0,1} \cos q_y 
-  2 \alpha_2 C_{0,1,1}) 
\; . 
\label{6}
\end{eqnarray} 
Note that our scheme preserves the rotational symmetry in spin space, i.e.\
$\chi^{zz} ({\bf q}, \omega ) \equiv \chi({\bf q},\omega)=\frac 12
\chi^{+-}({\bf q},\omega)$. For $|{\bf q}| \ll 1$ we have
\begin{equation}
\omega_{{\bf q}}^2 = c_x^2 q_x^2 + c_y^2 q_y^2 + c_z^2 q_z^2 \; ,
\label{7}
\end{equation}
with the squared spin-wave velocities
\begin{eqnarray}
c_x^2 &=& \frac 12 - 3 \alpha_1^x C_{1,0,0} + \alpha_2 C_{2,0,0} - 2 R_y 
(\alpha_1^x C_{1,0,0}-\alpha_2 C_{1,0,0} ) 
- 2 R_z (\alpha_1^x C_{1,0,0} - \alpha_2 C_{1,0,1} ) \; ,
\label{8} \\
c_y^2 &=& R_y^2 ( \frac 12 - 3 \alpha_1^y C_{0,1,0} + \alpha_2 C_{0,2,0}) 
- 2 R_y (\alpha_1^y C_{0,1,0} - \alpha_2 C_{1,1,0} )
- 2 R_y R_z (\alpha_1^y C_{0,1,0} - \alpha_2 C_{0,1,1} ) \; ,
\label{9} \\
\mbox{and} && \nonumber \\
c_z^2 &=& R_z^2 ( \frac 12 - 3 \alpha_1^z C_{0,0,1} + \alpha_2 C_{0,0,2}) 
- 2 R_z (\alpha_1^z C_{0,0,1} - \alpha_2 C_{1,0,1} )
- 2 R_y R_z (\alpha_1^z C_{0,0,1} - \alpha_2 C_{0,1,1} ) \; .
\label{10}
\end{eqnarray}

Considering the uniform static spin susceptibility $\chi = \lim_{{\bf q}\to 0}
M_{{\bf q}}^{(1)} / (2 \omega_{{\bf q}}^2 ) $, the ratio of the anisotropic
functions $M_{{\bf q}}^{(1)}$ and $\omega_{{\bf q}}^2$ must be isotropic in
the limit ${\bf q}\to 0$. That is, the conditions
\begin{equation}
(c_y/c_x)^2=R_yC_{0,1,0}/C_{1,0,0} 
\label{11}
\end{equation}
and
\begin{equation}
(c_z/c_x)^2=R_zC_{0,0,1}/C_{1,0,0} 
\label{12}
\end{equation}
have to be fulfilled.

The critical behavior of the model (\ref{1}) is reflected in our theory by the
closure of the spectrum gap at ${\bf Q}=(\pi,\pi,\pi)$ as $T$ approaches $T_N$
from above, so that $\lim_{T\to T_N} \chi^{-1}({\bf Q})=0$. At $T\leq T_N$ we
have $\omega_{{\bf Q}}=0$ and, separating the condensation part $C$, 
\begin{equation}
C_{{\bf r}}=\frac 1N \sum_{{\bf q}(\neq {\bf Q})} C_{{\bf q}} 
\mbox{e}^{ i {\bf q} {\bf r}} + C \mbox{e}^{ i {\bf Q} {\bf r}} \; ,
\label{13}
\end{equation}
where $C$ results from (\ref{13}) with ${\bf r}=0$ employing the sum rule
$C_{0,0,0}=\frac 12$. Then the staggered magnetization $m$ is calculated as 
\begin{equation}
m^2= \frac 1N \sum_{{\bf r}} \langle {\bf S}_0 {\bf S}_{{\bf r}} \rangle 
\mbox{e}^{- i {\bf Q} {\bf r}} = \frac 32 C 
\; .
\label{14}
\end{equation}
The theory has 14 quantities to be determined self-consistently (9 correlation
functions in $\omega_{{\bf q}}^2$, $m$, and 4 vertex parameters) and 13
self-consistency equations (10 Eqs.\ (\ref{13}) including $C_{0,0,0}=\frac
12$, the LRO condition $\omega_{{\bf Q}}=0$, and Eqs.\ (\ref{11}) and
(\ref{12})). If there is no LRO, we have $\omega_{{\bf Q}} > 0$, and the
number of quantities and equations is reduced by one. As an additional
condition for determining the free $\alpha$ parameter at $T = 0$, we adjust
the ground-state energy per site which we compose approximately as
$\epsilon(R_y,R_z)=\epsilon(R_y,0)+\epsilon(0,R_z)-\epsilon(0,0)$, where
$\epsilon(R_y,0)$ (and $\epsilon(0,R_z)$) is taken from the 
Ising-expansion results by Affleck et al.\ for the 2D spatially anisotropic
Heisenberg model, \cite{AGS94} and
$\epsilon(0,0)=-0.4431$ is the Bethe-ansatz value. This approximation is
suggested to 
be good at least for $R_z \ll R_y$ (or $R_y \ll R_z$). To get an additional
condition also at finite temperatures, where $\epsilon$ data are not available
and all vertex parameters are temperature dependent, we assume, following
Refs.\ \onlinecite{ST91} and \onlinecite{WI97}, the ratio
\begin{equation}
r_{\alpha}(T) \equiv 
\frac{\alpha_2(T)-1}{\alpha_1^x(T)-1} = r_{\alpha}(0)
\label{15}
\end{equation}
as temperature independent.

\section{Ground-state properties}

In Fig.~\ 1 our results for the zero-temperature staggered magnetization
$m_0\equiv m(T=0)$ as a function of $R_y$ and $R_z$ are shown. They indicate
an order-disorder transition at the phase boundary $R_{z,c}(R_y)$ or
$R_{y,c}(R_z)$ (cf.\ inset). For $R_z=0$ we get the critical ratio
$R_{y,c}(0)\simeq 0.24$ which was already found in Ref.~\
\onlinecite{ISW99}. In that paper the suppression of LRO below the finite 
value of $R_{y,c}$ was interpreted, in combination with ED data, as indication
of a rather sharp crossover in the spatial dependence of the spin correlation
functions in the LRO phase 
at the coupling ratio $R_{y,0}\simeq 0.2$. 
The finite value of $R_{y,c}$, however, seems to be due to the
approximations in our theory, since there are strong indications for
$R_{y,c}=0$ (see Ref.\  \onlinecite{Sch96}). 
Accordingly, we cannot explain the tiny magnetic
moments of Sr$_2$CuO$_3$ and Ca$_2$CuO$_3$, \cite{Koj97} since for $R_y
\ll 1$, \cite{REH97} we have $m=0$. This result is just opposite to the
overestimation of $m$ by all previous spin-wave theories. As seen in the phase
diagram (inset of Fig.~\ 1), the inclusion of the interplane coupling $R_z$
stabilizes the LRO, where this effect is quite considerable even at very small
values of $R_z$. 

Figure~\ 2 exhibits some short-ranged spin correlation functions at $T=0$. For
$R_z=0$, in  Ref.\ \onlinecite{ISW99} the correlators $C_{1,0,0}, C_{0,1,0}$, 
and $C_{1,1,0}$ as functions of $R_y$ were found to agree well with the ED 
data. For $R_z=0.02$ (cf.\ Fig.~\ 2) our results deviate only slightly from
those at $R_z=0$. The sign changes and magnitudes of $C_{\bf r}$ reflect the
AFM SRO. In the limit $R_y \to 0$ the correlations between the $x$-$z$-planes
vanish. At $R_z > R_{z,c}(0)\simeq 0.24$ the LRO enhances the inter-$x$-$z$
plane correlators and results in their sharp drop towards their limiting value 
$C \mbox{e}^{i \bf{Q} \bf{r}}$ as $R_y \to 0$. This is
visible in the data for $R_z=0.35$ in Fig.\ 2.

\section{Finite-temperature results}

At nonzero temperatures we have solved the self-consistency equations
(\ref{13}) supplemented by the conditions (\ref{11}), (\ref{12}), and
(\ref{15}) to obtain the magnetization $m(T)$, the N\'{e}el temperature
($m(T_N)=0$), the static spin susceptibility, and the anisotropic correlation
lengths.

In Fig.~\ 3 the N\'{e}el temperature is plotted as a function of $R_z$. For
$R_z=0$ we get $T_N=0$  (see Ref.\ \onlinecite{WI97}), in agreement with the
Mermin-Wagner 
theorem. The increase of $T_N$ with $R_z$ is governed by the intra-$x$-$y$
plane 
anisotropy. At a fixed value of $R_z$, the decrease of $T_N$ with decreasing
$R_y$ is in accordance with the reduced zero-temperature magnetization (cf.\ 
Fig.~\ 1). 
Comparing our results for $R_y=1$ with previous RPA/mean-field approaches (see
Table~\ I), we ascribe the reduction of $T_N$ as compared with  Refs.\
\onlinecite{MSS92,Kop92}, and \onlinecite{SI93} to the improved description of
SRO. That is, 
the LRO is suppressed in favor of a paramagnetic phase with pronounced AFM
SRO. If $R_z$ is fit to the N\'{e}el temperatures of real systems, the strong
overestimation of $T_N$ by previous theories results in very small values of
the interplane coupling. In our approach the resulting $R_z$ values turn out
to be higher. Considering La$_2$CuO$_4$ with $T_N=325$K  \cite{Kei92,Bir95}
and putting $J=130$meV ($J \equiv J_x=J_y$) or $J=117$meV, \cite{WI97} we
obtain 
$R_z\simeq 10^{-3}$ or $R_z \simeq 1.6\times 10^{-3}$, respectively, in
contrast to $R_z < 10^{-4}$ according to  Refs.\ \onlinecite{MSS92} and
\onlinecite{SI93}. For 
Ca$_{0.85}$Sr$_{0.15}$CuO$_2$ ($T_N = 540$K, $J=125$meV) \cite{MRB99} we get
$R_z\simeq 1.2 \times 10^{-2}$ as compared with $R_z\simeq 2.5\times10^{-2}$
obtained from a fit of the low-temperature magnetization data. \cite{MRB99}

Figure~\ 4 shows the temperature dependence of the staggered magnetization at 
$R_y=1$ (for the zero-temperature values, compare with Fig.~\ 1). The shape of
the normalized curve $m/m_0$ versus $T/T_N$ (see inset) depends on the single
parameter $R_z$ and is similar to that found in previous spin-wave theories. 
\cite{MSS92,STS92,Kop92,Liu90} At low enough temperatures the system
exhibits 3D behavior, so that the decrease of $m$ follows a $T^2$ law. This
was also observed by NMR experiments on La$_2$CuO$_4$  \cite{MMR97} ($T_N=
312$K) yielding $m/m_0=1 - a(T/T_N)^2$ with $a=0.67$ for $T\lesssim 100$K. The
NMR 
data is indicated in the inset of Fig.~\ 4 (marked by a bold curve) and agrees
well with our theory for $R_z = 10^{-3}$ (as estimated above). For
temperatures close to $T_N$ our 
numerical results for $m(T)$ are described by the law
$m(T) \propto (1-T/T_N)^{1/2}$. The square-root temperature behavior agrees
with 
the findings of  Refs.\ \onlinecite{STS92,Kop92,Liu90}, and with the
neutron scattering data on La$_2$CuO$_4$, \cite{Kei92} but contradicts the
result of  Ref.\ \onlinecite{IKK99} ($m \propto 1-T/T_N$).

Considering the AFM correlation lengths above $T_N$ and for $R_y=1$, the
expansion of $\chi({\bf q})$ around ${\bf Q}$, 
$\chi({\bf q}) = \chi({\bf Q}) \left[1+\xi^2_{xy}(k^2_x+k^2_{y}) 
+\xi^2_z k^2_z \right]^{-1}$ with ${\bf k}={\bf q} - 
{\bf Q}$, yields the intraplane correlation length
\begin{eqnarray}
\xi^2_{xy}&=&-\omega^{-2}_{{\bf Q}}\left[\frac{1}{2} + 11\alpha^x_1 C_{1,0,0} +
\alpha_2\left(C_{2,0,0} + 2C_{1,1,0}\right)+\right.\nonumber\\
&&\left.+2R_z\left(\alpha^x_1 C_{1,0,0} +2\alpha^z_1 C_{0,0,1} + \alpha_2
C_{1,0,1}\right)\right] -\frac{2C_{1,0,0}}{M_{{\bf Q}}^{(1)}}
\label{16}
\end{eqnarray}
and the interplane correlation length
\begin{eqnarray}
\xi^2_z&=&-R_z\omega^{-2}_{\bf Q}\left[4\left(2\alpha^x_1 C_{1,0,0} +
\alpha^z_1 C_{0,0,1} + \alpha_2 C_{1,0,1}\right)+\right.\nonumber\\
&&\left.+R_z\left(\frac{1}{2}+5\alpha^z_1 C_{0,0,1} + \alpha_2 C_{0,0,2} 
\right)\right]-\frac{2R_zC_{0,0,1}}{M_{\bf Q}^{(1)}} \; . 
\label{17}
\end{eqnarray} 
In Fig.~\ 5 the influence of the interplane coupling on the temperature
dependence of $\xi^{-1}_{xy}$ and $\xi^{-1}_z$ (inset) is shown.
For comparison, the intraplane correlation length at $R_z=0$  (see also Ref.\
\onlinecite{WI97}) is 
plotted, where the low-temperature expansion
$\xi_{xy}=2(2\alpha^x_1|C_{1,0,0}(0)|)^{1/2}T^{-1}
\exp[2\pi\alpha^x_1m_0^2/(3T)]$ holds up to $T=0.2$ within a deviation of
about $6 \%$ from the full temperature dependence calculated by
Eq.(\ref{16}). For $R_z > 0$ the correlation lengths diverge at $T_N$, since
the gap $\omega_{\bf Q}$ closes as $T$ approaches $T_N$ from above. In the
vicinity of $T_N$, $\xi^{-1}_{xy}$ and $\xi^{-1}_z$ behave as $T-T_N$ also
found by previous mean-field approaches. \cite{SI93,IKK99}

Let us compare our results for  the intraplane correlation length with the
neutron-scattering data on La$_2$CuO$_4$  \cite{Bir95} in the range 340K$\le T
\leq 820$K shown in Fig.\ 6. Taking $J$ as obtained previously  \cite{WI97}
from a least-squares fit of $\xi_{xy}$ in the 2D model ($a=3.79$\AA), 
$J=117$meV, for $T > 500$K and $R_z\lesssim 3.5\times 10^{-3}$
we get a good quantitative agreement with experiments. In  Ref.\
\onlinecite{WI97} the 
deviation of the theory for $R_z=0$ and $T < 500$K from the experimental data
was ascribed to the appearance of the preexponential factor $T^{-1}$ in the
low-temperature expansion of $\xi_{xy}$ which is an artifact of our mean-field
approach. However, this deviation may be reduced by the inclusion of the
interplane coupling, since $\xi^{-1}_{xy}(T_N)=0$. For $T_N=325$K
\cite{Bir95} we get $R_z \simeq 1.6\times 10^{-3}$ (see above, Fig.\ 3), and
the theoretical $\xi^{-1}_{xy}$ curve lies between the $R_z=0$ result and the
experiments. The discrepancy between the theoretical and experimental
low-temperature correlation lengths may be further reduced by the choice of
higher $R_z$ values. Taking, for example, $R_z=3.4\times10^{-3}$, we get a
very good quantitative agreement (cf.\ Fig.\ 6) down to 360K; however, the
N\'{e}el 
temperature turns out to be somewhat higher ($T_N=353$K).

Finally, we consider the uniform static spin susceptibility $\chi(T)=
\lim_{\bf q\to 0}\chi(\bf q)$. In Fig.~\ 7 the anisotropy effects on the
temperature dependence are demonstrated. For $R_z=0$ and a strong intraplane
anisotropy ($R_y < 0.2$) the minimum of $\chi(T)$ at a finite temperature,
being an artifact of our approach, may signal the crossover in the spatial
dependence of the spin correlation functions at $R_{y,0}\simeq 0.2$ as was
discussed in Sec.~\ III. Note that such a minimum in the 1D model ($R_y=0$) 
was also found in  Ref.\ \onlinecite{KY72}. At $R_y > 0.2$, the increase of
$\chi$ with temperature, the maximum at $T_{max}$ near the exchange
energy $J_x=1$ (see inset), and the crossover to the high-temperature
Curie-Weiss behavior are due to  the decrease of AFM SRO with increasing
temperature (cf.\  Ref.\ \onlinecite{WI97}). Let us point out that the
susceptibility maximum is totally missed in RPA theories. \cite{STS92} With
increasing $R_y$, we obtain an increase of $T_{max}$ which agrees
with a general tendency found in various spin--$1/2$ Heisenberg models and
analyzed in  Ref.\ \onlinecite{JohR}. For comparison, the exact values at
$R_y =0$ and $R_y=1$ are given by $T_{max}=0.64$  \cite{BF64} and
$T_{max}=0.94$, \cite{OK88} respectively. Since our theory allows 
the calculation of $T_{max}$ at any spatial anisotropy, it may provide
a reliable interpretation of experimental data on low-dimensional spin
systems. Considering the maximum spin susceptibility
$\chi_{max}=\chi(T_{max})$, again our results are in accordance with
the general behavior:  \cite{JohR} $\chi_{max}$ increases with
decreasing $T_{max}$, i.e. with decreasing $R_y$. Concerning the
influence of the interplane coupling, the enhancement of the low-temperature
susceptibility by $R_z$ may be explained by the weakening of the SRO effect in
higher dimensions. As seen from Figs.\ 7 and 3, the uniform susceptibility has
no singularity at the N\'{e}el temperature, contrary to the RPA result of 
Ref.\ \onlinecite{STS92} revealing a peak of $\chi(T)$ at $T_N$. Concerning the
maximum in $\chi(T)$ of La$_2$CuO$_4$, we get $T_{max}=1.19J=1615$K
(cf.\ Fig.\ 7, $J=117$meV). This value roughly agrees with the estimate given
by Johnston,  \cite{Joh89} $T_{max}=1460K$, by means of a scaling
analysis of the susceptibility data below 800K.

\section{Summary}

In this paper we have extended the spin-rotation-invariant Green's-function
theory of magnetic LRO and SRO in 2D Heisenberg models  \cite{WI97,ISW99}
to the 3D Heisenberg antiferromagnet with arbitrary spatial anisotropy. Our
theory provides a satisfactory interpolation between the low-temperature and
high-temperature behavior, where the temperature dependent SRO, described in
term of two-spin correlation functions, is adequately taken into account. The
main results are summarized as follows.

{\bf (i)}\ \ The incorporation of SRO results in a strong suppression of
N\'{e}el order with increasing anisotropy and in a reduced N\'{e}el
temperature as compared with previous spin-wave approaches.

{\bf (ii)}\ \ The temperature dependence of the uniform static spin
susceptibility reveals 
a maximum in the short-range ordered paramagnetic phase and a crossover to
the Curie-Weiss law. The position of the maximum is influenced by the spatial 
anisotropy.

{\bf (iii)}\ \ Comparing the theory with experiments on the magnetization and
correlation length of La$_2$CuO$_4$, a good quantitative agreement is found.

From the results of our theory we conclude that the application of this
approach to extended Heisenberg models (anisotropy in spin space, frustration)
may be promising to describe the SRO effects on the unconventional magnetic
properties of real low-dimensional spin systems.\\[20mm]

Acknowledgments: The authors, especially L.\ Siurakshina, are very grateful
to the DFG for financial support.  Additional support by the Max-Planck
society and the INTAS organisation
(INTAS-97-1106) is acknowledged. The authors thank S.-L.\ Drechsler for many
useful discussions.

\newpage

{\bf TABLE I.}\ \ \ N\'{e}el temperature $T_N/J_x$ at $R_y=1$ compared with
other approaches\\[7mm] 
\begin{center}
\begin{tabular}{||c|c|c|c|c||}
\hline\hline
$-\log_{10}(R_z)$ & Fig.~\  3   &Ref.\ \onlinecite{MSS92}  
&Ref.\ \onlinecite{Kop92} &Ref.\ \onlinecite{SI93}\\[1mm]
\hline
4 &   0.17  & 0.48  & 0.29  & 0.25\\[1mm]
3 &   0.22  & 0.65  & 0.38  & 0.34\\[1mm]
2 &   0.36  & 0.80  & 0.54  & 0.47\\[1mm]
1 &   0.56  & 1.15  &       & 0.68\\[2mm]
\hline
\end{tabular}\\[15mm]
\end{center}

\begin{center}
{\Large\bf Figures}
\end{center}

Fig.\ 1. Staggered magnetization at $T=0$ as a function of spatial
anisotropy. The inset shows the stability region of N\'{e}el order.

Fig.\ 2. Spin correlation functions at $T=0$ for different spatial
anisotropies. 

Fig.\ 3. N\'{e}el temperature as a function of $R_z=J_z/J_x$. 

Fig.\ 4. Staggered magnetization vs.\ temperature for $R_y=1$. The inset shows
the $R_z$ dependence of the normalized curves compared with the NMR data on
La$_2$CuO$_4$ \cite{MMR97} (bold curve).
 
Fig.\ 5. Inverse antiferromagnetic  correlation lengths within
($\xi^{-1}_{xy}$) and between the $x$-$y$ planes ($\xi^{-1}_z$, see inset) for
$R_y=1$.  

Fig.\ 6. Inverse antiferromagnetic intraplane correlation length in
La$_2$CuO$_4$ obtained by the neutron-scattering experiments of  Ref.\
\onlinecite{Bir95}
and
from the theory ($R_y=1$) for different $R_z$ values. 

Fig.\ 7. Uniform static spin susceptibility vs.\ $T$. The inset exhibits the
position $T_{max}$ of the maximum in $\chi(T)$ vs.\ $R_y$.

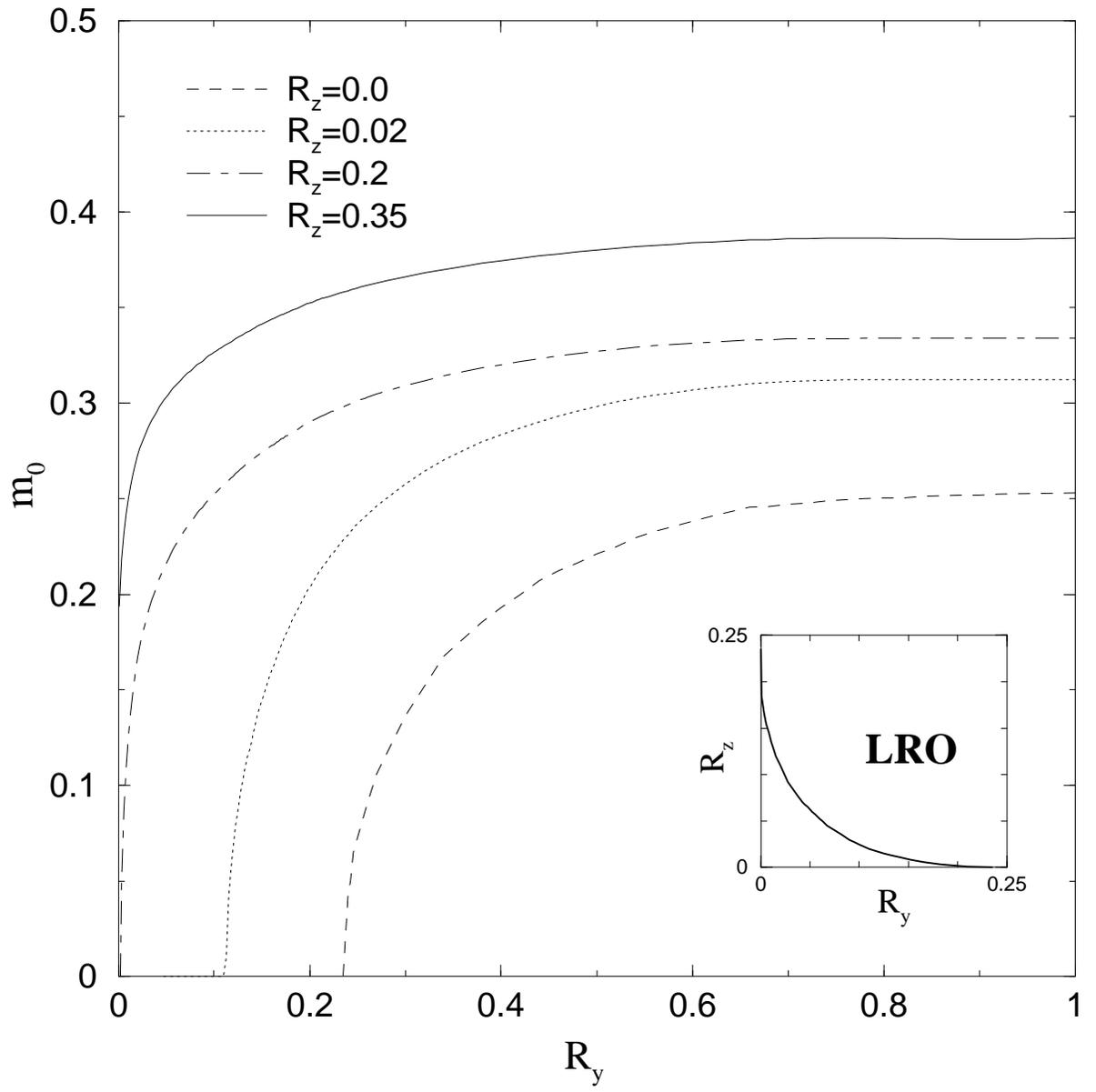
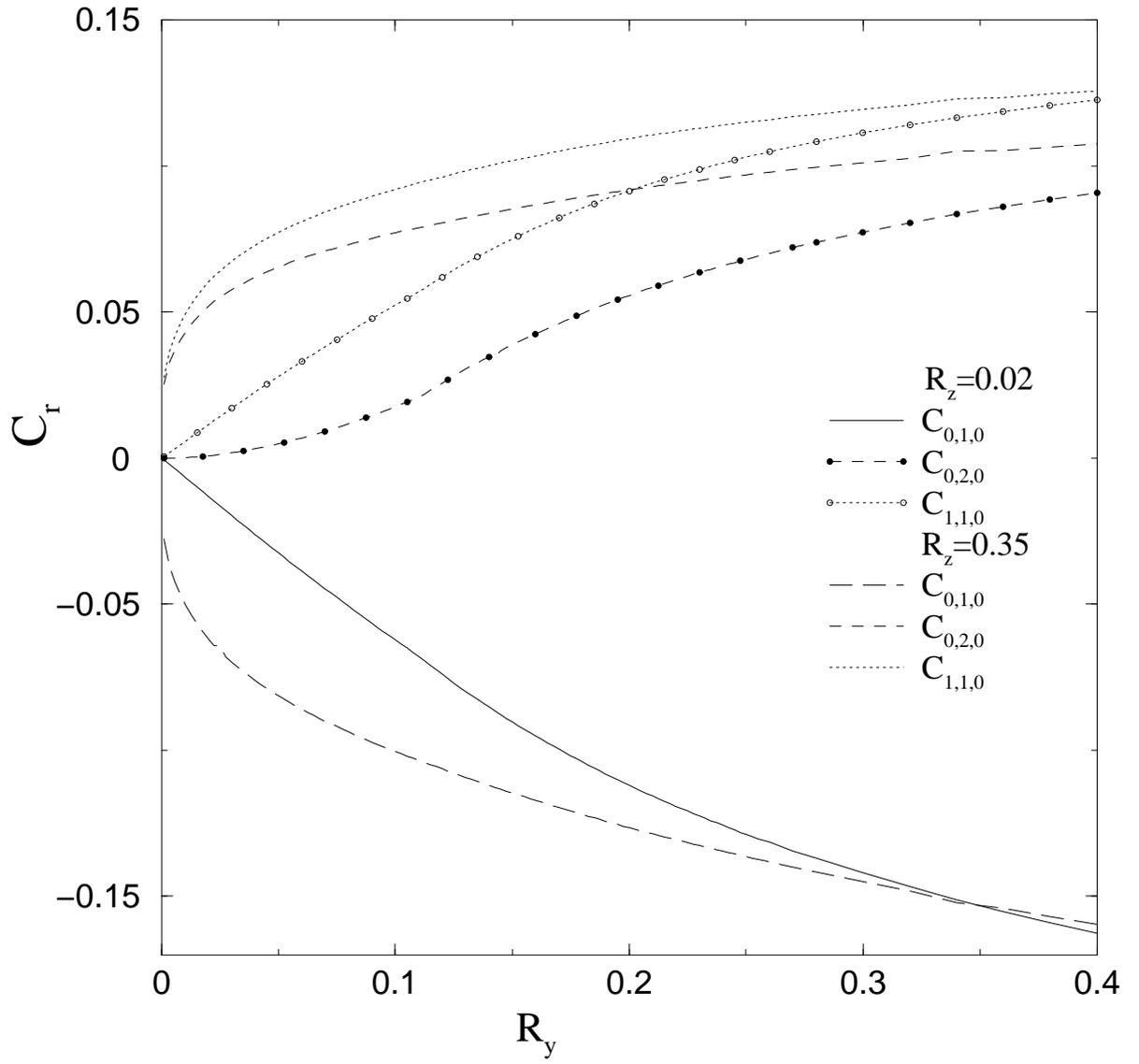
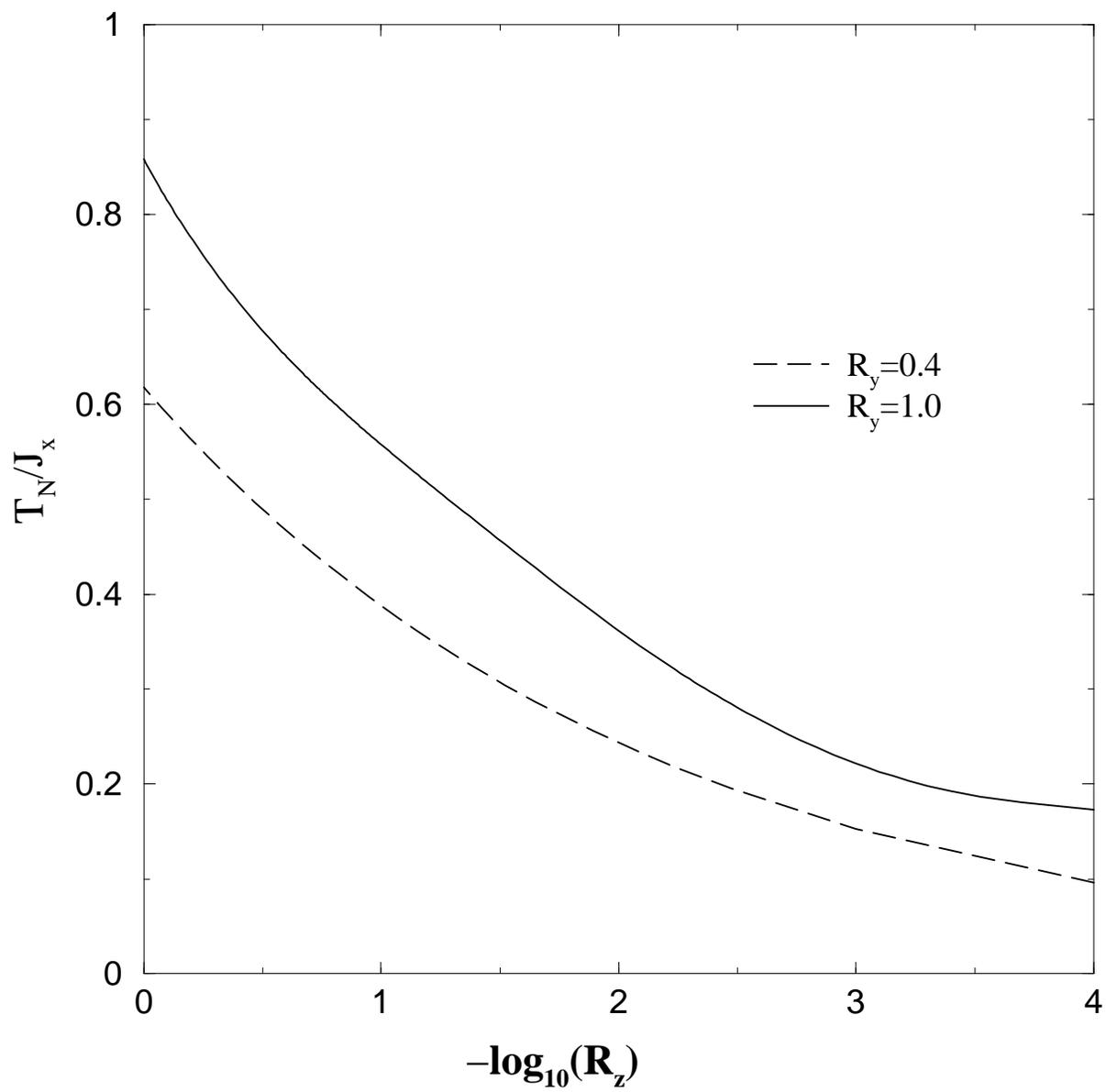
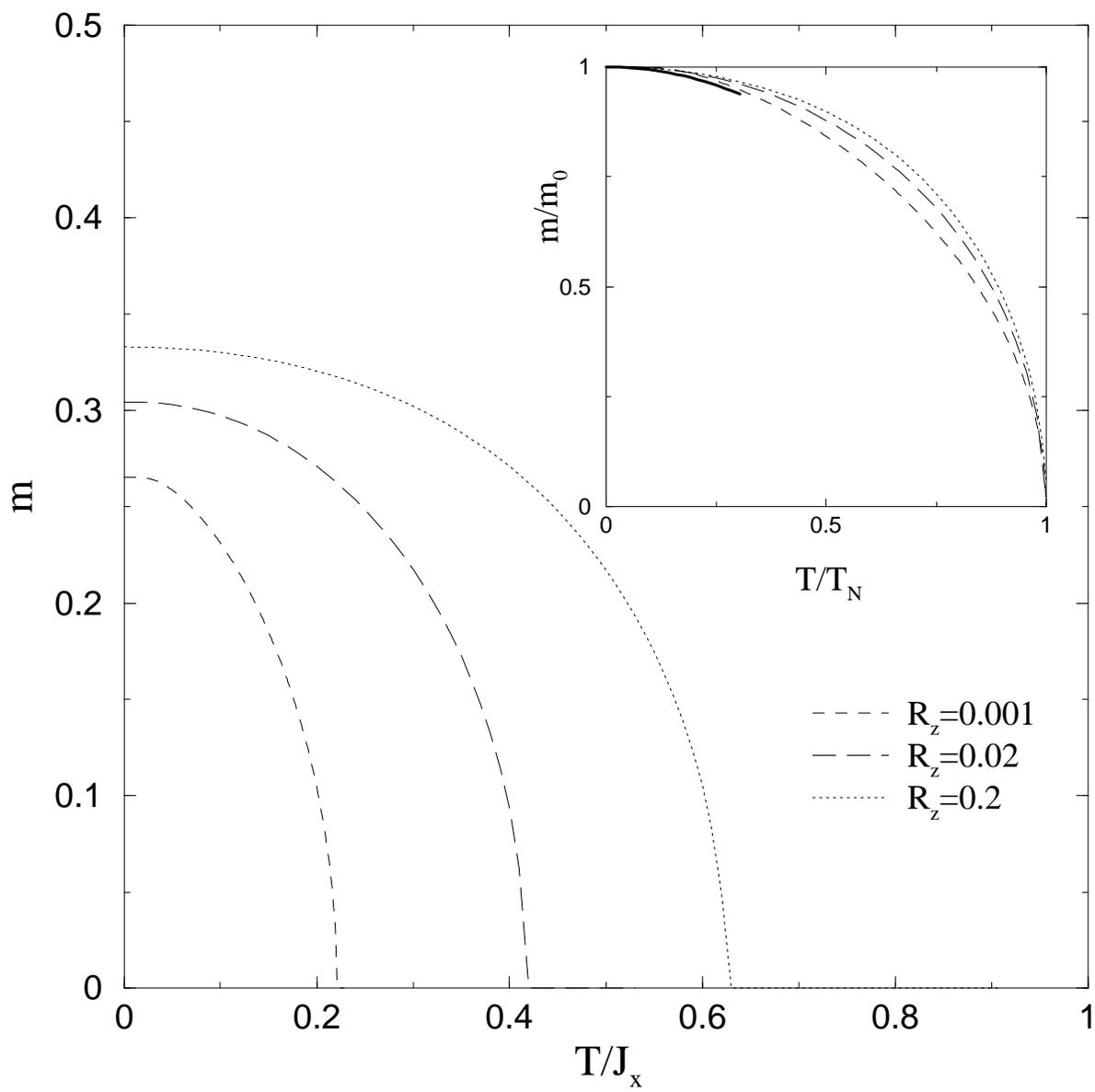
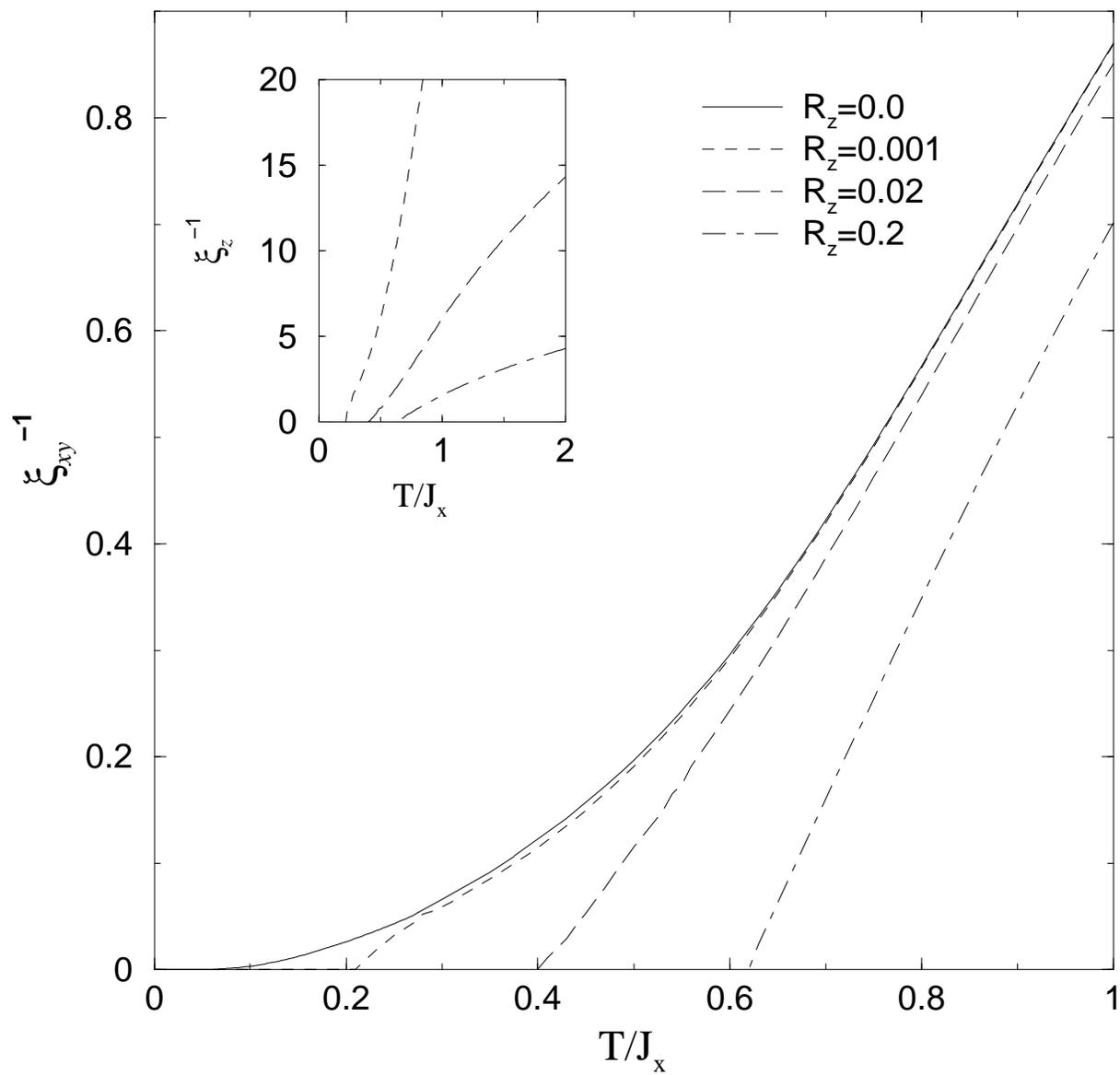
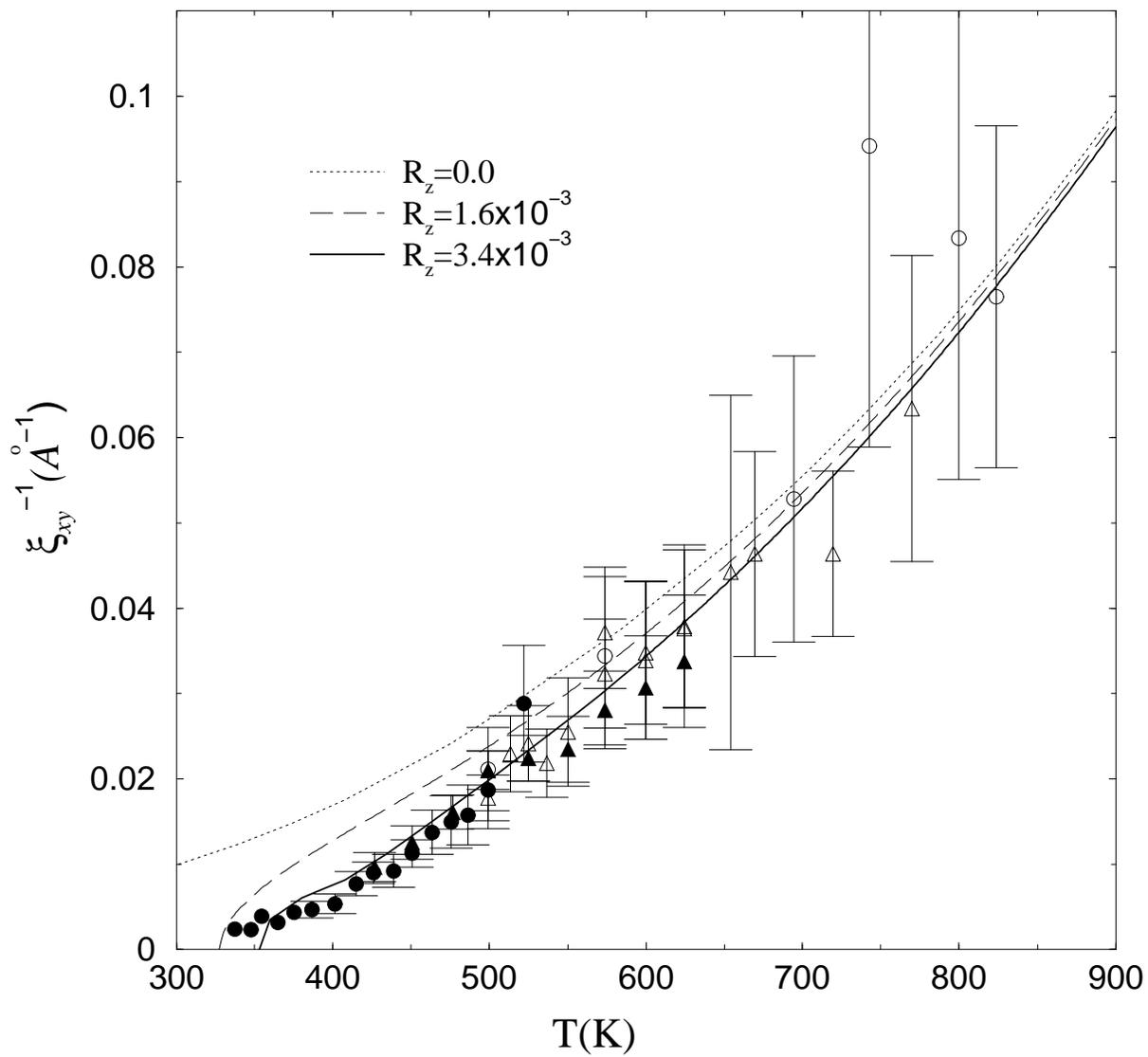
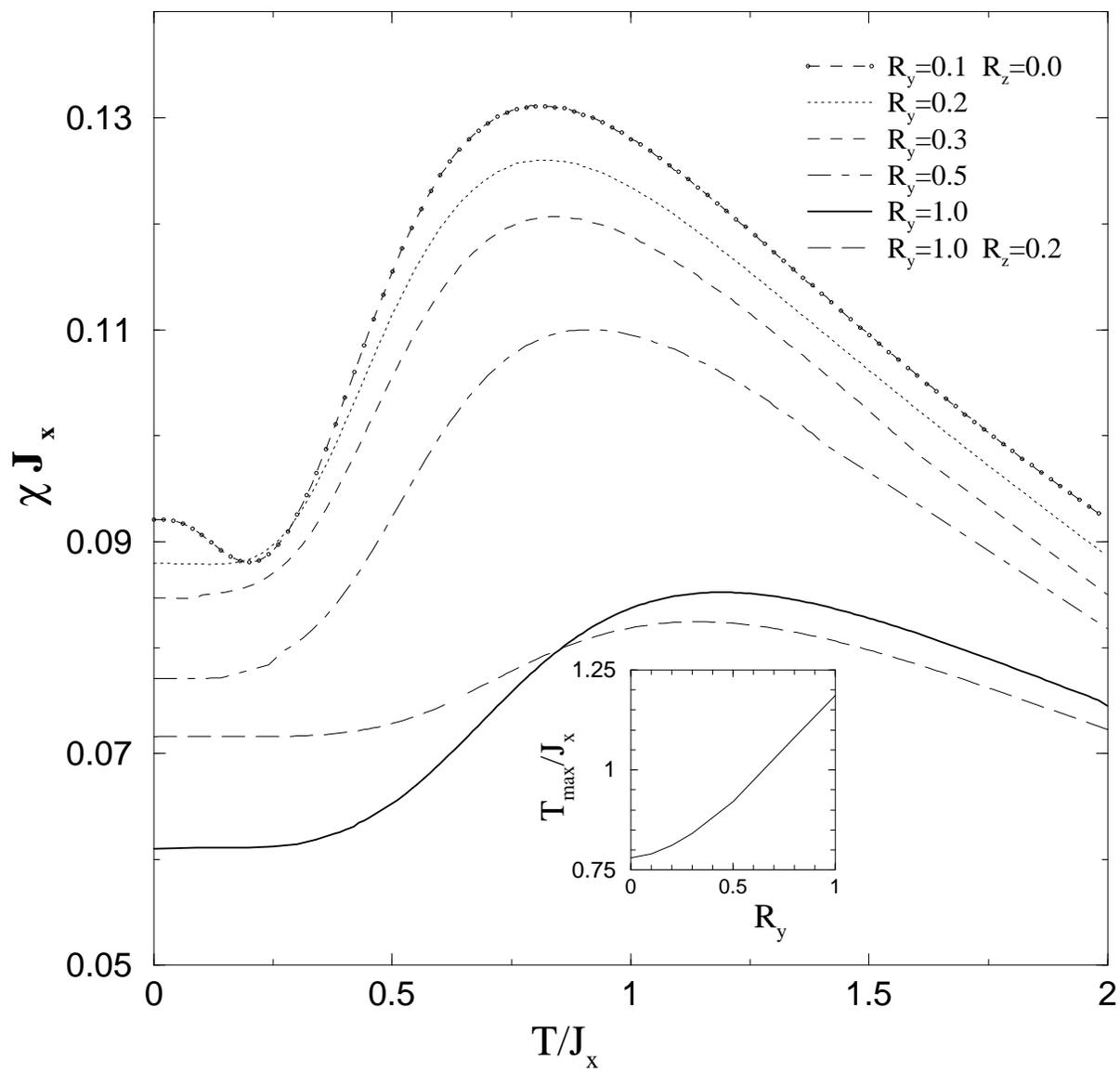
\end{document}